# Effects of the Functional Group on the Lithium Ions Across the Port of Carbon Nanotube

Xie Hui[†], Luo Geng, Yang Cheng-Bing, Luo Chong, and Liu Chao

(Key Laboratory of Low-grade Energy Utilization Technologies and Systems of Ministry of Education, College of Power Engineering, Chongqing University, Chongqing, 400030, China)

## Abstract

The mean axial velocity of lithium irons across the entrance of carbon nanotube $V_{Li}$ is an important factor for the charge-discharge performances of rechargeable Lithium battery. The molecular dynamics simulation method is adopted to evaluate the factors and their effects on $V_{Li}$ which include the diameter of carbon nanotube, functional group type on the port and the number of a given type of functional group. The statistical analysis of the calculation results shows that: In the selected carbon nanotubes of four different diameters, $V_{Li}$ will gradually rise with the increase of CNT diameter due to lithium irons migration resistance decreasing; as the port of CNT is successively modified to hydrogen (-H), hydroxyl (-OH), amino ($-NH_2$) and carboxyl (-COOH), the corresponding migration resistance of lithium ions is enhanced resulting in the dropping of $V_{Li}$; in comparison to the effect strength of four types of functional groups on $V_{Li}$, -COOH shows strongest, $-NH_2$ and -OH perform relatively weaker, and the effect difference between $-NH_2$ and -OH is very small, -H displays weakest; When the number of a given non-hydrogen functional group on the port sequentially increases, it also shows a trend that lithium ion migration resistance gradually increases which makes $V_{Li}$ decreases in turn. The more influential the functional group, the greater the impact of functional group number changes on $V_{Li}$. The results of this paper have some significance on the precise production of lithium-ion battery electrode materials, enhancing the overall battery cycle efficiency and charging speed.

*This work is supported by the National Natural Science Foundation of China (Grant No. 51206195), Natural Science Foundation of Chongqing (Grant No. cstc2013jcyjA90009) and the Fundamental Research Funds for the Central Universities of Ministry of Education of China (Grant No. CDJZR12110033).
[†]Corresponding author. Email: xiehui@cqu.edu.cn ; Tel: .023-65102469

# 1. Introduction

Nowadays, energy power battery has become the key points and the main bottleneck of a product in the process of its innovation and development when facing a fierce and competitive technical developing environment[1-3]. Especially for the products known to the public such as computers, mobile phones, electric vehicles and wireless charging technologies, possessing high mass-energy ratio, rapid charging and discharging rate, low price, safety and environmental friendly battery has become an important competitive advantage[4-6]. Amongst various energy and power batteries, rechargeable Li-ion batteries (LIB) are receiving more and more researches and applications in the field of new energy power because of its high excellent energy density and longer cycle life[1]. As the performances of Li-ion batteries are often restricted to its component materials, researching and designing excellent electrode materials and developing electrolyte of high efficient transmission performance have been the central themes of plentiful scientific researches to promote their properties[6, 7]. CNT (Carbon Nano Tube), with its unique structure, excellent electrical conductivity, strong mechanical strength and chemical stability[8-11], has been used to construct electrode materials for Li-ion batteries, which can greatly strengthen the charge-discharge characteristic and cycle performance[12]. Comparing to historically using graphite or carbon black as anode material for LIBs, combining CNTs with anode can greatly improve the battery performance as a consequence of its appropriate structure for lithium iron intercalation and diffusion[13].

In order to further investigating the mechanism of lithium iron migrating in and out of the CNT, many scholars have done lots of researches. Zhao et al.[14] found it was accessible for lithium atom to embed both inside and outside of the CNT and Senami et al.[15] discovered the internal points of CNT were easier for lithium atom to embed by using the first principle method. The morphological structure of CNT will produce great influence on electrochemical performance of LIB. CNTs with different chiral structure will have an impact on the diffusion capacity of lithium ions. Kawasaki et al.[16] observed the reversible Li ion storage capacity of metallic CNT is about 5 times greater than that of semiconducting one by means of electrochemical charge-discharge measurements. The lateral defects on the surface of CNTs can increase the probability of lithium ions insetting into the CNT internal space on account of the migration resistance reducing. The larger the defects, the more easily for the diffusion of lithium iron[17]. Similarly, opening the ends of CNT can greatly improve the spreading performance of the lithium ion[18]. The enhanced capacity is attributed to lithium ions diffusion into the interior of the SWNTs through the opened ends and sidewall defects[19]. Experiments also were done to confirm that the LIBs with defective CNTs had a higher reversible capacity, initial columbic efficiency and a lower charge-transfer resistance[20]. Different lengths of CNT also have a very significant influence on the diffusion of lithium ions. Short CNTs can facilitate easier intercalation and deintercalation of lithium ions because the lithium ions are able to enter, but seldom exit if the tube is too long[21]. Radius is other geometric parameter of CNT affecting the lithium ion adsorption and diffusion. Zhang et al.[22] found that when used as anode materials for LIB, CNTs with different radius could influence the electrochemical performances in different degrees in his recent experimental studies.

In view of the importance of the morphological structure of CNT when used in LIB, many methods have been developed to create the desired CNTs, such as arc-discharge[23, 24], laser ablation[25], gas-phase catalytic growth from carbon monoxide[26] and chemical vapor deposition[27]. Commonly these processes can be used not only to create defects in CNT structures, cut CNTs into shorter lengths and remove the caps of CNTs, but also add a large amount of functional groups on the places where chemical bonds broken, such as hydrogen (-H), hydroxyl (-OH), amino (-$NH_2$) and (-COOH) group[28]. The topological structures and electronic cloud distribution of these functional groups will affect the lithium ion migration velocity, which can lead to substantial voltage hysteresis. In order to further investigate how the morphology of carbon nanotubes and functional groups affect the capacity of lithium ion diffusion, we will use molecular dynamic simulation method to study the effect on the mean axial velocity of lithium irons across the entrance of carbon nanotube $V_{Li}$, with different diameter of CNTs and the different type of functional groups modified around the port of CNT.

## 2. Simulation details

2.1 The simulation model

Considering the existing research found that the irons could pass through the corresponding CNTs when the diameter was just greater than 10 Å[29], we performed the simulation by building four types of armchair single-walled carbon nanotubes with each diameter respectively being 12.20Å (9, 9), 14.92 Å (11, 11), 17.63 Å (13, 13), 20.34 Å (15, 15) and the length being 39.35 Å. Four types of functional group (hydrogen -H, hydroxyl -OH, amino -$NH_2$ and carboxyl -COOH) were fully added at positions spread uniformly around the top and bottom rim of the CNT. The simulation system considered a hexagonal box with dimensions 20.98×20.98×58.00 Å (Fig.1). The CNT was located in the middle of the box with its axis paralleling to Z axis and the distance from the top and bottom of the box to CNT respectively was 10 Å. We made the CNT solvate into lithium chloride aqueous solution of 5.6mol/L (LiCl) and the periodic condition was applied in all directions. With CNT (13, 13) as an example, the system is shown in Fig.1.

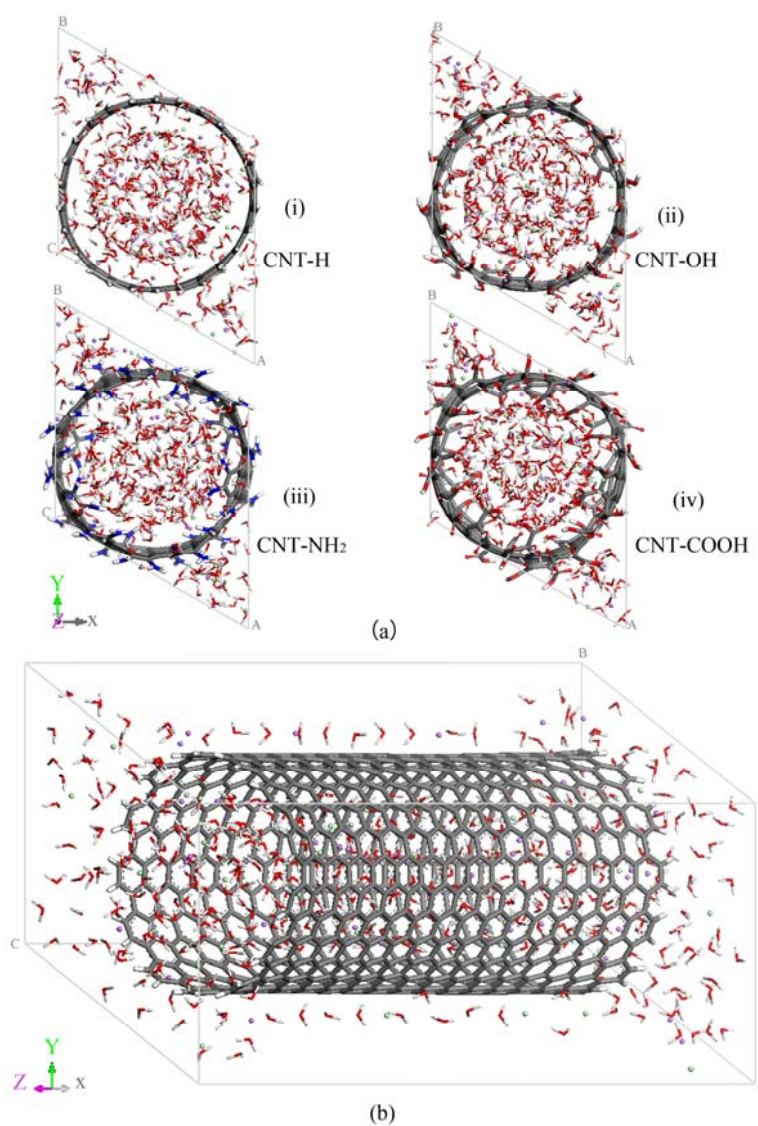

**Fig.1** The initial simulation system. The purple and orange balls represent lithium and chlorine irons, respectively, while the red, blue and grey sticks denote oxygen, ammonia and carbon atoms. (a) Top view. (i) ~ (iv) respectively depict the situation that four types of functional groups (hydrogen -H, hydroxyl -OH, amino -NH$_2$ and carboxyl -COOH) were fully added at positions spread uniformly around the top and bottom rim of the CNT. (b) Stereogram view.

In addition, in order to explore the effect on $V_{Li}$ with a change of a given functional group number,

we respectively modified the number of four types of functional group around the top and bottom rim of the CNT. The research object is based on CNTs (13, 13), of which the functional groups added on the ends separately consist of a combination of -H and -OH or -H and -NH$_2$, -H and -COOH. In each combination, the number of non-hydrogen functional group increases from 6 to 26 with 4 as tolerance. Considering the combination of -H and -NH$_2$ in Fig.2, the number of -NH$_2$ in each CNT respectively is 6, 10, 14, 18, 22, and 26 with the corresponding number of -H being 20, 16, 12, 8, 4, and 0. It is similar in the other two situations.

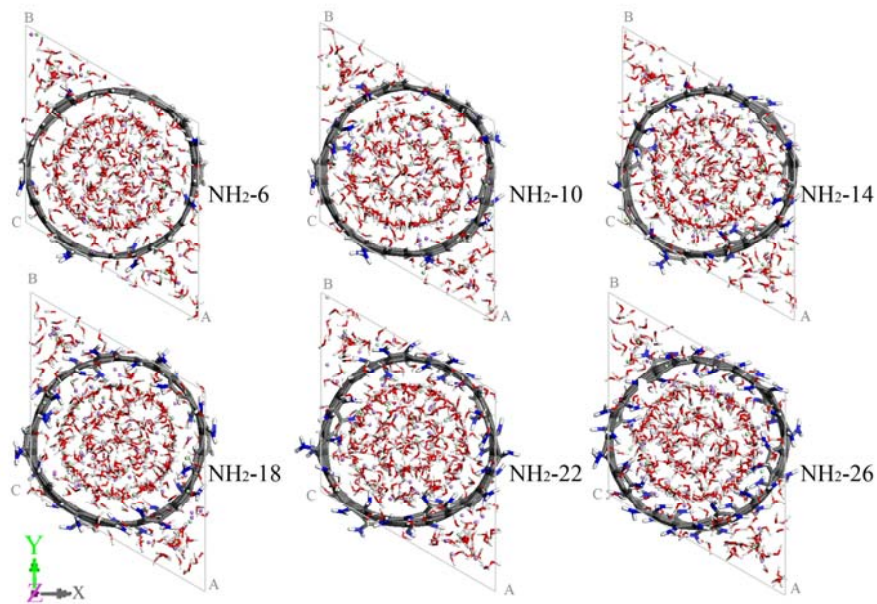

**Fig.2** The functional groups of -H and -NH$_2$ at the ends of CNT

Furthermore, this paper studied the effect on the mean axial velocity of lithium irons across the entrance of carbon nanotube $V_{Li}$, by changing the diameter of CNT and the type of functional group added around the ends of CNT. The velocity was separately calculated in the region S$_1$, S$_2$, S$_3$ and S$_4$ (Fig.3). The ranges of influence of four regions are discrepant duo to the lengths of molecular chains are various in different functional groups. So we defined a length difference of 0.53 Å along z axis among S$_1$, S$_2$, S$_3$, and S$_4$, respectively corresponding to CNT modified by hydrogen (CNT-H), by hydroxyl (CNT-OH), by amino (CNT-NH$_2$) and by carboxyl (CNT-COOH).

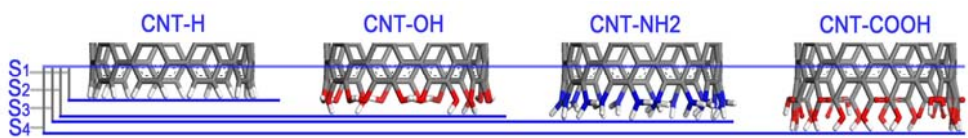

**Fig.3** The region calculated for $V_{Li}$ in four types of CNT

2.2 The simulation process

The COMPASS force field was applied in the simulation, which is based on *ab initio* force field that enables accurate simulation. All systems were energy minimized and equilibrated, and the molecular dynamic simulations were performed with NVT ensemble at 295 K and a time step of 1 fs. The temperature was controlled by the Andersen method and Coulomb interaction was calculated using Ewald summation method. At the initial stage of the simulation, the system was running to reach a given state of thermal equilibrium. When the systems were running about 0.1ns and the speeds of the particles were in line with Maxwell-Boltzmann distribution, then the system could be considered to have reached equilibrium. Then the systems were applied an external electric field of 1.0V/ Å, which is in the range of research strength[30], and continued to run for 1ns. We collected data after the systems again reached a steady state.

## 3. Results and discussion

In order to verify the reasonableness of the model, we firstly analyzed the radial and axial density distribution of water molecules in CNT in equilibrium without electric field. With CNT (13, 13) as an example, the results are shown in Fig.4. The radial density distribution of water molecules is described in Fig.4 (a), with x axis representing the radius of CNT and y axis the dimensionless local water density. As can be seen from the Fg.4 (a), a layered structure of water is produced in carbon nanotubes, and has a prominent peak near the pipe wall. Due to the impact of the surface potential energy, the fluid molecules appear a certain orderly distribution of layered and spatial fluctuations near the wall region of nanochannel[31]. This phenomenon has been observed in numerous literatures[32-35]. Fig.4 (b) reveals the axial density distribution of water molecules, which consistent with the distribution of water molecules along the CNT in the LiCl aqueous as described in the literature[36].

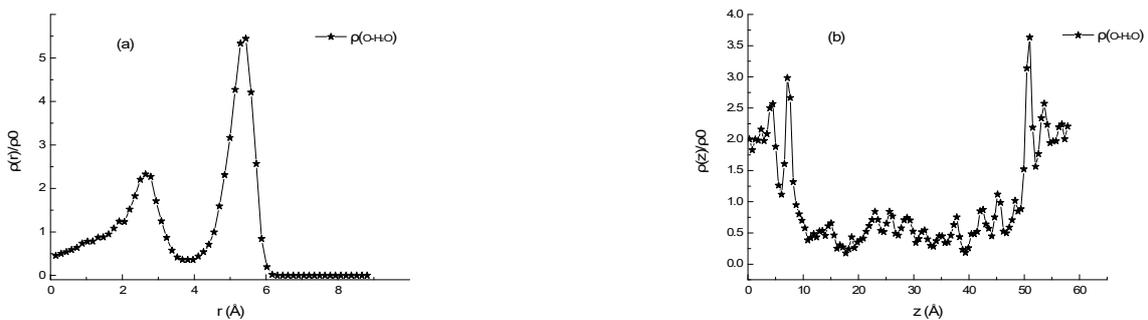

**Fig.4** Density profiles of the oxygen atoms of water molecules in CNT (13, 13) with oxygen atom of water molecule being used to present the whole water molecular. $\rho(r)/\rho_0$ is the dimensionless local water density, where $\rho(r)$ is the actual and $\rho_0$ the total mean density. (a) Radial density profiles, and x-axis = 0 corresponds to the center of the CNT; (b) Axial density profiles.

3.1 Influence of the CNT diameter and the types of functional group on the port

The distribution trends of $V_{Li}$ under the influence of different CNT diameters and the different types of functional group on the port of each given diameter CNT are depicted in the Fig.5. Firstly, we focus on the effects of CNT diameter on $V_{Li}$. Within the given scales, we prepared four different diameters which respectively were CNT (09, 09), CNT (11, 11), CNT (13, 13) and CNT (15, 15). As can be seen from the figure, no matter which type of functional group was added on the port of CNT, $V_{Li}$ showed a trend of gradual increase with the diameter of CNT by Dr=12.20 Å (09, 09) gradually increasing to Dr=20.34 Å (15, 15). When the rim of CNT port was fully modified into -COOH, lithium ion axial mean velocity by $V_{Li}$ =4.75 Å /Ps gradually increased to 12.25 Å /Ps with the increase of CNT diameter, nearly three times the rate of increase. When the port ring of CNT was fully modified into -H, lithium ion axial mean velocity by $V_{Li}$ =4.75 Å /Ps when Dr=12.20 Å, gradually increased to 12.25 Å /Ps when Dr=12.20 Å, nearly an increase of 4.21 Å /Ps. Similarly, when the port of CNT was fully modified into -OH or -NH$_2$, $V_{Li}$ still increased with the increase of CNT diameter, despite the small increase in amplitude. When the -OH was fully added on the port of CNT, $V_{Li}$ possessed an increase of 1.49Å/Ps and when the port fully modified into -NH$_2$, then $V_{Li}$ had an increase of 3.1Å/Ps. Obviously, different CNT diameters affect the speed of lithium ions moving through the port. That is, in the studied range, increasing the diameter of CNT can facilitate the insertion of lithium irons. This is consistent with the research results of Li Hongman et al.[36] which found that the permeability of lithium ions increased gradually with the diameter increasing from CNT (09, 09) to CNT (11, 11), when they studied the effects of different CNT diameters on the migration of lithium irons. We can understand this phenomenon from two aspects. On the one hand, increasing the diameter of the CNT enlarges effective flow area of lithium irons, reducing its migration resistance at the port of CNT. On the other hand, the Van Der Waals forces and Coulomb forces between lithium irons and carbon atoms of CNT become weaker with the increase of CNT diameter, which reduces the migration resistance at the port so that increases the insertion velocity.

Then we turn our attention to the effects of different types of functional group on the port of each given diameter CNT. As shown in Fig.5, when the port of CNT was fully modified into -H, $V_{Li}$ was relatively at a larger level showed by $\overline{V_{Li}}$=26.58 Å/Ps in the CNTs of four different diameters. When the port was all modified into -OH or -NH$_2$, $\overline{V_{Li}}$ slowed little difference respectively corresponding to 15.38, 14.92 Å/Ps and the distribution curves of velocity were also close. When the port was fully modified into -COOH $\overline{V_{Li}}$ was reduced to 8.88Å/Ps. In the CNTs of diameter $Dr$=12.20Å, as the port of CNT was successively fully modified into -H, -OH, -NH$_2$ and -COOH, $V_{Li}$ gradually decreased which respectively were 23.47, 14.99, 13.38, 4.75 Å/Ps. In the CNTs of the other three diameters $Dr$=14.92, 17.63, 20.34 Å, the corresponding $V_{Li}$ had a similar reduction distribution trend. As can be found from the statistical results, $V_{Li}$ shows obvious difference in the CNTs of the same diameter but different types of functional group added on the port. This indicates that the CNTs with its port modified into different functional groups have a certain degree of influence on lithium migration. In addition, because of the effects of -OH and -NH$_2$ on $V_{Li}$ are close, the values of $V_{Li}$ in the two kinds of modified CNT were similar and the velocity distribution curves and even have alternation phenomenon. Thus in comparison to the effect strength of four types of functional groups on $V_{Li}$, -COOH shows strongest, -NH$_2$ and -OH perform relatively weaker, and the effect difference between -NH$_2$ and -OH is very small, -H displays weakest. In the research of Thepsuparungsikul N et al., battery electrode material based on CNT with -OH provided better results than that based on CNT with -COOH[37], which is similar with the results calculated above.

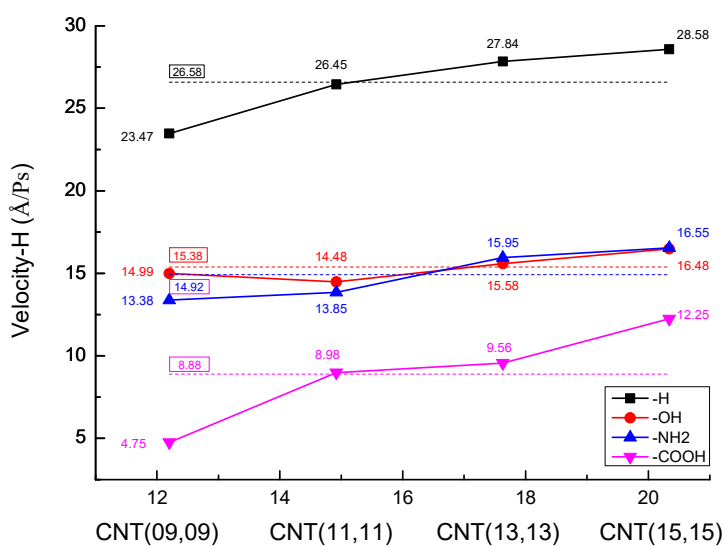

**Fig.5** The distribution trends of $V_{Li}$ under the influence of different CNT diameters and the different types of functional groups on the port of each given CNT diameter. The diameters respectively are CNT (09, 09), CNT (11,

11), CNT (13, 13), CNT (15, 15), corresponding the x-axis and the types of functional groups are –H, -OH, -NH$_2$, -COOH which are separately described by the color black, red, blue and pink. The solid line represents the actual velocity and the dotted line represents the mean speed of four different $V_{Li}$s along the x-axis which is represented by $\overline{V_{Li}}$.

3.2 Influence of the number change of a given type of functional group on the port

We have learned from the results calculated above that the CNTs with its port modified into different types of functional group have a certain degree of influence on lithium migration. In order to further investigate the effects of functional group, this paper also explores the effects on $V_{Li}$ with a change of a given functional group number. We respectively modified the number of four types of functional group around the top and bottom rim of the CNT. The research object is based on CNT (13, 13), of which the functional groups added on the ends separately consist of a combination of -H and -OH or -H and -NH$_2$, -H and -COOH. In each combination, the number of non-hydrogen functional group increases from 6 to 26 with 4 as tolerance as shown in Fig.2.

We calculated and collected data of $V_{Li}$ after the system reached a stable state at the same condition with T=298K and E=1.0V/Å. The results were shown in Fig.6. Regardless of the differences of functional group type on the port, $V_{Li}$ showed a trend of gradual decreasing and the extent of reduction performs fast firstly then low, with the number of a given non-hydrogen functional group increasing on the port. When the non-hydrogen functional group becomes -COOH, the overall velocity of lithium ions were lower than the ones that functional group is -OH or -NH$_2$. When the number of functional group -COOH increasing from 0 to 26 until the carbon ring bonds saturated on the port, the mean value of $V_{Li}$, namely $\overline{V_{Li}}$, equals to 17.71 Å/Ps. Replacing the -COOH with -OH or -NH$_2$ to do the same modified variation, the statistical average value $\overline{V_{Li}}$ respectively corresponds to 20.51, 20.37 Å/Ps. Similarly, the CNTs with its port modified into different functional groups have a certain degree of influence on migration velocity of lithium irons, which is consistent with the previous statistical results. Among three types of non-hydrogen functional group, -COOH showed the strongest influence. Not only did it make $V_{Li}$ relatively at the level of a lower rate, but also made $V_{Li}$ decrease with its number increase and the extent of reduction the largest. The effects of -OH and -NH$_2$ on $V_{Li}$ are very close. When the port ring of CNT was respectively modified into these two kinds of functional group, $V_{Li}$ also decreased with the corresponding functional group number increase, but $V_{Li}$ was relatively at the level of a higher rate.

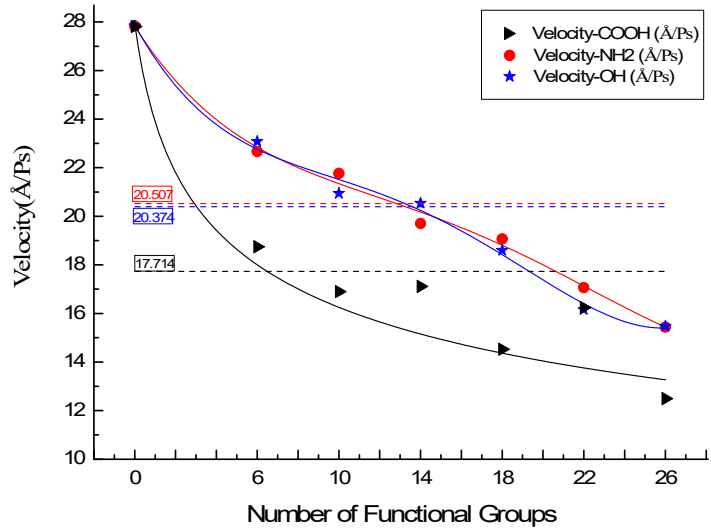

**Fig.6** The distribution trends of $V_{Li}$ under the influence of different types and number of functional groups on the port. The types of functional groups are -OH, -NH$_2$, -COOH which are separately described by the color blue, red, and black. The solid line represents the actual velocity and the dotted line represents the mean speed of four different $V_{Li}$s along the x-axis which is represented by $\overline{V_{Li}}$.

Thus, the influence of the types of functional group and its number on $V_{Li}$ can be summarized as follows: As the port carbon ring is successively modified to -H, -OH, -NH$_2$ and -COOH, the corresponding migration resistance of lithium ions is enhanced resulting in the dropping of $V_{Li}$; in comparison to the effect strength of four types of functional groups on $V_{Li}$, -COOH shows strongest, -NH$_2$ and -OH perform relatively weaker, and the effect difference between -NH$_2$ and -OH is very small, -H displays weakest; When the number of a given non-hydrogen functional group on the port sequentially increases, it also shows a trend that lithium ion migration resistance gradually increases which makes $V_{Li}$ decreases in turn. The more influential the functional group, the greater the impact of functional group number changes on $V_{Li}$.

Lithium ions will be affected by the space resistance of the other ions or atoms of functional groups on the port of CNT and the particles absorbed on these functional groups when inserting into the CNT or the gaps between the CNT tubes. When the port carbon ring is successively modified into -H, -OH, -NH$_2$ and -COOH, the number of atoms in functional groups increases accordingly so that the scopes of its influence become larger. And then the effective flow area of lithium irons through the CNT will reduce, which makes the migration resistance increase and the velocity decrease around the port. In the same way, when the number of non-hydrogen functional group gradually increases, the number of atoms in functional groups will also increase, making the extent of influence stronger.

In order to verify the explanation that the number of functional group affects the effective flow area of lithium irons through the CNT, We performed the simulation to gain the radial density profiles of the lithium irons in CNT (13, 13) with the number of functional groups increasing from 0 to 26 on the condition of T=298K, E=1.0V/Å, as shown in Fig.7. The number of functional group does have a certain degree of influence on migration velocity of lithium irons. When the port was fully modified into -H without -COOH, the intensity of resistance showed relatively weakest and the amount of lithium ions into the CNT was most, making the relative density the biggest. With the -H constantly replaced by -COOH, lithium ions into CNT decreased gradually and the relative density then reduced little by little. Until the port was fully modified into -COOH, the lithium ions into the CNT reached a minimum, the relative density also decreased to a minimum. This shows that the number of functional groups on the port do affect the lithium ions into the carbon nanotubes within the effective flow area, which affects the lithium ion radial density distribution in the carbon nanotubes.

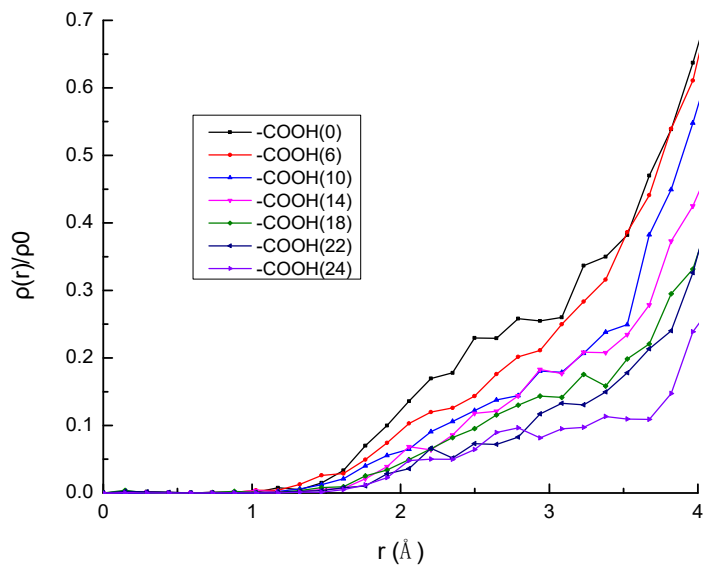

**Fig.7** Radial density profiles of the lithium irons in CNT (13, 13) with the number of functional group -COOH increasing from 0 to 26. $\rho(r)/\rho_0$ is the dimensionless local density, where $\rho(r)$ is the actual and $\rho_0$ the total mean density. And x-axis = 0 corresponds to the wall of the CNT.

The functional groups added in this paper are polar molecules. There is a strong electrostatic interaction between them and the lithium ions. This can be similarly deduced from the phenomenon that water molecules have preferential dipole orientations pointing toward the -COOH groups[38]. Ben

Corry[39] found that water and ion transport through CNTs functionalized by different charged and polar functional groups on the port would suffer from an obvious resistant barrier. This is consistent with the research results of the influence that polar functional groups added on the entrance of CNT have on the migration of lithium ions in this paper. Compare to the weak Van der Waals forces between carbon atoms of CNT and the lithium ions, adding polar functional groups on the port will cause strong electrostatic interaction between lithium ion and the polar functional group atoms. Lithium ions will be attracted to the functional group atoms in the process of entering the port of CNT, showing a significant impediment. In the four types of functional group, the resistance caused by -H is the weakest so that the lithium ions can enter the CNT with greater speed. While the -OH and -NH$_2$ not only have a similar size of molecular, but also possess a similar amount of charge, so the resistances that these two kinds of functional group exhibit are close but stronger than the one added by -H. The -COOH on the entrance of CNT has the longest branch and thus the polar oxygen atoms of its have a larger range of motion. So the influence of Coulomb forces and the Van der Waals forces between the atoms of -COOH and lithium ions thereby can be extended, which makes the resistance on the port become the strongest and the lithium ions enter the CNT with the slowest velocity. In the same way, with the number of the non-hydrogen functional group on the entrance of CNT increase, the forces caused by polar atoms and charge will gradually accumulate, resulting in the increase of lithium ions migration resistance. This is also similar to the results that the strength of water molecular attracted by -COOH will become stronger as the number of -COOH increases[38].

## 4. Conclusion

The factors including the diameter of carbon nanotube, functional group type on the port and the number change of a given type of functional group, have been chosen to study their effects on the mean axial velocity of lithium irons across the entrance of carbon nanotube $V_{Li}$, by using the method of molecular dynamic simulation. The conclusion can be drawn as follows: In the selected four different diameter of carbon nanotube, $V_{Li}$ will rise gradually with the increase of CNT diameter due to lithium irons migration resistance decreasing; as the port of CNT is successively modified to hydrogen (-H), hydroxyl (-OH), amino group (-NH$_2$) and carboxyl group (-COOH), the corresponding migration resistance of lithium ions is enhanced resulting in the dropping of $V_{Li}$; in comparison to the effect strength of four types of functional groups on $V_{Li}$, -COOH shows strongest, -NH$_2$ and -OH perform relatively weaker, and the effect difference between -NH$_2$ and -OH is very small, -H displays weakest; When the number of a given non-hydrogen functional group on the port sequentially increases, it also shows a trend that lithium ion migration resistance gradually increases which makes $V_{Li}$ decreases in turn. The more influential the functional group, the greater the impact of functional group number changes on $V_{Li}$. The results of this paper have some significance on the precise production of lithium-ion battery electrode materials, enhancing the overall battery cycle efficiency and charging speed.